\documentclass[a4paper,twocolumn,english,conference]{IEEEtran}
\usepackage[utf8]{inputenc}
\usepackage{amsmath,amssymb,amsfonts}
\usepackage{array,multirow,graphicx}
\usepackage{textcomp}
\usepackage{xcolor}
\usepackage{float,caption,subcaption}
\usepackage{enumitem}
\usepackage{algorithm,algpseudocode,setspace}
\usepackage{gensymb}
\begin{document}

\title{\vspace{-0.2cm}Proactive Scheduling and Caching for Wireless VR Viewport Streaming\vspace{-0.2cm}}

\author{\IEEEauthorblockN{
Mostafa Abdelrahman\IEEEauthorrefmark{1},
Mohammed Elbamby\IEEEauthorrefmark{2}
and Vilho R{\"a}is{\"a}nen\IEEEauthorrefmark{2}
}
\IEEEauthorblockA{
\IEEEauthorrefmark{1}Aalto University, Espoo, Finland mostafa.abdelrahman@aalto.fi\\
\IEEEauthorrefmark{2}Nokia Bell Labs, Espoo, Finland \{mohammed.elbamby, vilho.raisanen\}@nokia-bell-labs.com
\vspace{-0.35cm}}}

\maketitle

\begin{abstract}
Virtual Reality (VR) applications require high data rate for a high-quality immersive experience, in addition to low latency to avoid dizziness and motion sickness. One of the key wireless VR challenges is providing seamless connectivity and meeting the stringent latency and bandwidth requirements. This work proposes a proactive wireless VR system that utilizes information about the user's future orientation for proactive scheduling and caching. This is achieved by leveraging deep neural networks to predict users' orientation trained on a real dataset. The 360$\degree$ scene is then partitioned using an overlapping viewports scheme so that only portions of the scene covered by the users' perceptive field-of-view are streamed. Furthermore, to minimize the backhaul latency, popular viewports are cached at the edge cloud based on spatial popularity profiles.  Through extensive simulations,  we show that the proposed system provides significant latency and throughput performance improvement, especially in fluctuating channels and heavy load conditions. The proactive scheduling enabled by the combination of  machine learning prediction and the proposed viewport scheme reduces the mean latency by more than 80\% while achieving successful delivery rate close to 100\%.
\end{abstract}


\vspace{-0.2cm}\section{Introduction}
\label{chapter:intro}

Over the last decade, virtual reality (VR) has seen significant advancements to improve the rendering capabilities and provide a more immersive experience \cite{mandal_brief_2013}. This has gained a lot of attention from businesses and consumers and paved the way for use cases such as interactive video games and 360$\degree$ videos \cite{elbamby_toward_2018}.  However, many if not all, of today's high-end VR headsets require a wired connection to a powerful computer to provide intensive rendering computation and user tracking. The setup cost for a powerful PC and the reduced mobility due to wired connections are amongst the main barriers to mainstream adoption of such headsets. On the other hand, some existing mid-range headsets provide better movement flexibility by utilizing local on-device computational power. These headsets however provide limited capabilities and also dissipate heat as the load increases, which is not favorable as the head-mounted display (HMD) is close to the eyes. 

An alternative approach is to rely on wireless connectivity to cloud servers to carry out the heavy rendering while the VR headset acts as a thin client. Remote centralized clouds will induce high network and communication latency, which is not suitable due to the latency sensitivity of VR applications. Studies \cite{potter_detecting_2014,albert_latency_2017} have shown that a latency of more than 15ms in VR can lead to motion sickness. Increased latency also breaks the immersion of the VR experience. Moreover, VR applications have high bandwidth requirements as content is projected inside a sphere and transmitting the entire 360$\degree$ view at each time frame is often not feasible on a wireless connection. Fortunately, only a limited region of the sphere is within users' field-of-view (FoV). Still, high bandwidth is required to transmit only the FoV. The study in \cite{elbamby_toward_2018} estimated a required bit rate of 1 Gbps for a 150x120$\degree$ FoV. Therefore, wireless VR headsets can provide a better experience, but various optimizations need to be implemented to reduce the latency and not restrict VR to simple and low-resolution applications. 

%\subsection{Related Work}

 Streaming only the active FoV of a 360$\degree$ video requires having prior knowledge about the users' FoV of the individual frames. Much work has been done to predict the users' FoV for future frames. Prediction methods can be classified into trajectory-based, content-based and hybrid methods. Trajectory-based methods utilize historical head movements to extrapolate future orientation. For example, the authors in \cite{bao_shooting_2016} developed a neural network regression model to predict the user's viewport based on the roll, pitch and yaw trajectories. In \cite{petrangeli_trajectory-based_2018}, spectral clustering was used to cluster trajectories and an individual trajectory is extrapolated based on the cluster average. The work in \cite{ban_cub360_2018} used a linear regressor to predict an individual user future point, the prediction is then amended via a K nearest neighbor election method of nearest users. However, trajectory-based methods can suffer from huge drops in accuracy for large prediction horizons ($>$ 2 seconds).  Content-based methods, such as in \cite{xu_predicting_2018} and \cite{li_very_2019}, extract saliency features from the video frame to predict future regions of interest. As for hybrid methods, the saliency features and trajectory information are combined. Examples are the studies in \cite{rondon_track_2020} and \cite{nguyen_your_2018}. 

From wireless streaming perspective, recent works have investigated incorporating machine learning (ML) prediction into optimizing streaming quality in wireless networks. For example, the studies in \cite{qian_flare_2018,qian_optimizing_2016} use FoV prediction to determine relevant tiles to improve bandwidth consumption, whereas in \cite{bao_motion-prediction-based_2017}, a transmission scheme incorporating both multicast and unicast is proposed to maximize bandwidth efficiency. The work of \cite{perfecto_taming_2020} used a recurrent neural network to directly predict future FoV tiles of the 360$\degree$ frame, and the prediction is used to implement a proactive multicast transmission scheme in a multi-user scenario. Similarly, other works have focused on viewport-based streaming methods in which pre-partitioned and rendered portions of video frames are streamed to users. For instance, \cite{bao_shooting_2016} used two deep neural networks to predict the viewpoint center and the possible deviation in the viewport center prediction, both predictions are used to calculate and adaptive viewport size that contains the FoV with high probability. Joint optimizing of VR transmission and caching is studied in \cite{Mingzhe_VR_caching}. However, ML is used for network management optimization but not for VR FoV prediction.

%\subsection{Problem statement}

This paper proposes proactive wireless VR streaming method. The proposed method predicts user's future viewport and subsequently assigns wireless resources to users and caches popular viewports in a proactive manner such that stringent low latency requirements are met.  More concretely, the contributions of the paper are summarized as follows:
\begin{itemize}[leftmargin=.0cm,itemindent=0.5cm,labelwidth=\itemindent,labelsep=0cm,align=left]
\item The method uses a deep neural network that processes the time series of orientations to predict users' orientation for various time horizons to minimize latency as users navigate the virtual environment.
\item The 360$\degree$ scene sphere is partitioned into small segments, called viewports, and only the relevant segments are fetched. Viewport-based rendering is shown to provide robustness against non-perfect predictions, i.e., less powerful models could be used without significant reduction in overall performance. 
\item Viewport-based rendering also enables offline analysis and caching of popular video viewports in edge clouds based on spatial popularity profiles, resulting in improved bandwidth usage, reduced backhaul latency and fewer quality transitions. 
\item The performance of the proposed approach is demonstrated using extensive simulations. It is shown that the proposed machine learning method achieves a mean angular error less than 15$\degree$ for a prediction horizon up to 30 frames. Proactive network scheduling  combined with overlapping viewport scheme are shown to not only reduce the mean latency by more than 80\% but also to achieve a success delivery rate close to 100\%, particularly in fluctuating channels. Caching of popular viewports at the edge cloud is shown to improve the 99-percentile delay by an additional 10\%. 
 
\end{itemize}
Th rest of this paper is organized as follows. The overall model and architecture of our proposed system is presented in Section \ref{chapter:system}. The implementation of machine learning prediction and network optimization components are then described in Section \ref{chapter:implementation}. The experimental setup and analysis of the results are detailed in Section \ref{chapter:evaluation}. Finally, conclusions are drawn in the last Section. 
 
\section{System Model}
\label{chapter:system}
\begin{figure}[htbp]
  \begin{center}
    \includegraphics[clip,trim=9cm 6cm 8cm 3cm,height=5cm,width=0.9\columnwidth]{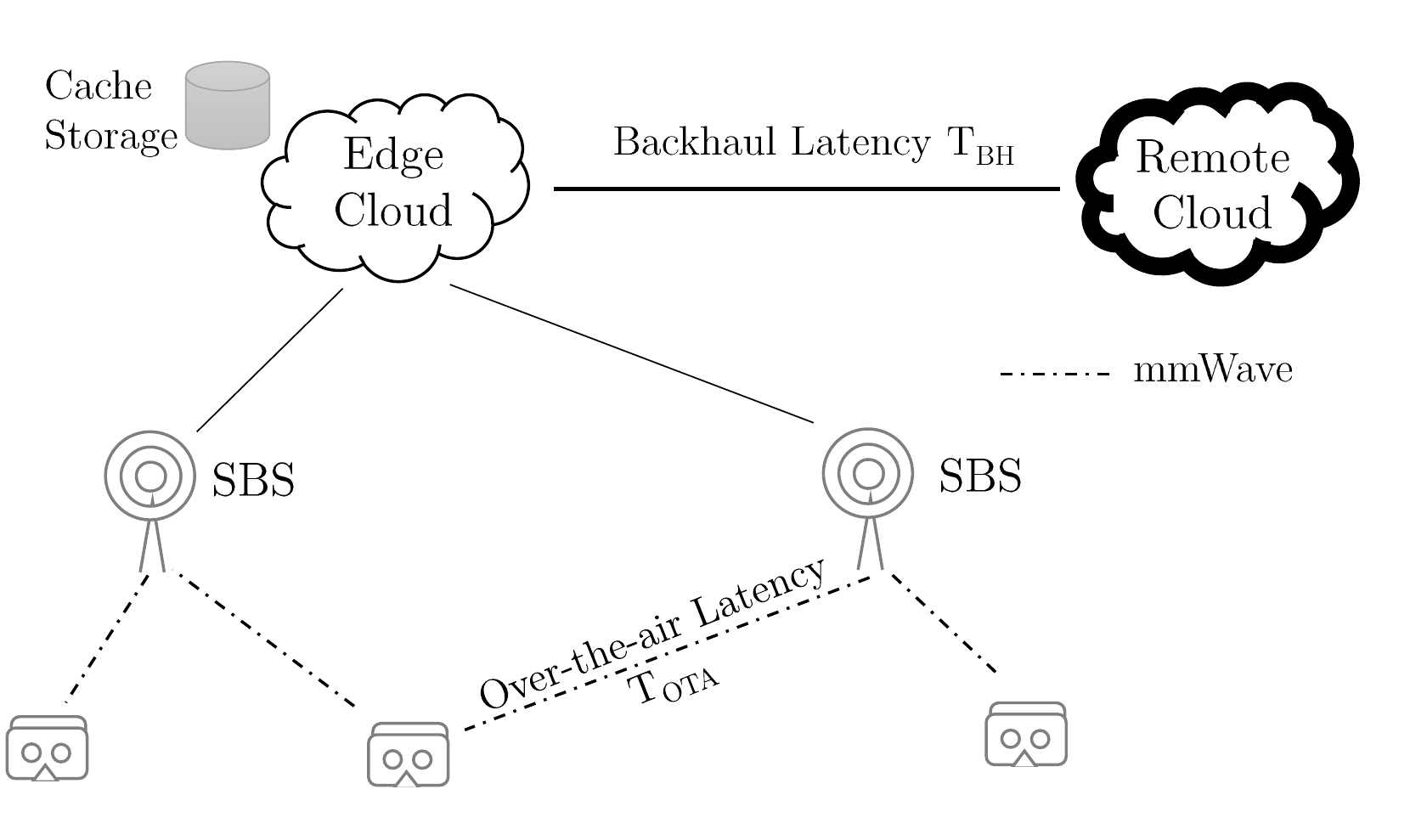}
    \caption{Illustration of network architecture for indoor VR arcade with mmWave SBSs and edge cloud computing.}
    \label{fig:network_architecture}
  \end{center}
\end{figure}

Consider an indoor VR arcade with dimensions $l \times w$ and a set of users $\mathcal{U}$, uniformly distributed throughout the arcade, each equipped with millimeter wave (mmWave) HMDs. Each user chooses to watch a 360$\degree$ video $v$ from the available video catalog $\mathcal{V}$. The network consists of multiple mmWave small base stations (SBS) $\mathcal{B}$ at fixed locations distributed around the arcade. These SBSs are connected to an edge cloud server equipped with a cache storage. While the user is watching a particular video, the user location and 3DoF pose are continuously tracked, uploaded and mapped into a set of candidate viewports to be streamed back to the users. An illustration of the network model is shown in Fig. \ref{fig:network_architecture}.

\subsection{Edge Architecture Model}

Due to the limited computational power and storage capacity of the HMDs, content is rendered in the remote cloud server and transmitted to the edge cloud with backhaul delay $T_{BH}$. The content is then transmitted via the SBSs to the users within the indoor arcade with over-the-air (OTA) latency $T_{OTA}$.  To minimize the backhaul latency, content is cached at the edge cloud. Because of limited edge storage capacity, only the most popular viewports are cached. For any video in the catalog, not all viewports are equally likely to be requested by users. Furthermore, the popularity changes with the content of the video i.e., a viewport popularity is a function of current frame index. Hence, the spatial popularity profile of any video is defined as a ranking of the viewports within a frame.
\subsection{Network Model}
\label{section:chapter3_network_model}
In mmWave communications, transmissions  are challenging due to sensitivity to blockage and high path loss in non-line-of-sight (NLOS) conditions \cite{niu_survey_2015,wei_key_2014}. Hence, mmWave can achieve higher bandwidth over a smaller denser area and it is well suited for the proposed VR arcade. We employ a probabilistic model that weights both line-of-sight (LOS) and NLOS path losses with the probability of LOS between a user $u$ and base station $b$. The path loss $\ell_{ub}$ can be written as:
\begin{equation}
\ell_{ub}(d) = p_t(d) \ell_{ub}^{\text{LOS}}(d) + (1-p_t(d)) \ell_{ub}^{\text{NLOS}}(d)
\end{equation}
where $p_t(d)$ is the LOS probability, $\ell_{ub}^{\text{LOS}}$ and $\ell_{ub}^{\text{NLOS}}$ are the LOS and NLOS path loss respectively and $d$ is the distance in meters between the SBS and the user.

For a user $u$, SBS $b$ and transmit power $p_b$, the Signal to Interference and Noise Ratio (SINR) is calculated as:
\begin{equation}
\label{eqn:sinr}
\text{SINR}_{bu}(t)= \frac{p_b G_c^{bu}(t) G_{Rx}^{bu}(t) G_{Tx}^{bu}(t)}{I_u(t) + \text{BW}_b N_0}
\end{equation}
where $G_c^{bu}(t)$, $G_{Rx}^{bu}(t)$, $G_{Tx}^{bu}(t)$ are the channel, receive and transmit antenna gains between user $u$ and SBS $b$, respectively $\text{BW}_b$ is the bandwidth for SBS $b$, $N_0$ is the noise power spectral density and $I_u(t)$ is the interference at time instant $t$.

\section{Proposed Approach}
\label{chapter:implementation}

Because of the low latency requirements of VR applications, we propose a joint prediction, matching and caching scheme. First, we utilize machine learning to predict users' head orientations for future frames.  Subsequently, the orientations are mapped into viewports that can be pre-rendered and proactively streamed to the user. Finally,  users are matched to SBSs and  popular viewports are cached.

\subsection{Pose Prediction}

Let $P_t$ denote the vectors of the 3DoF pose at time $t$ represented as a quaternion $q=w+x\boldsymbol{\hat{\textbf{i}}}+y\boldsymbol{\hat{\textbf{j}}}+z\boldsymbol{\hat{\textbf{k}}}$  and let $T_H$ be the prediction horizon.  Also, let $P_{t-t_p:t}=\{P_i\}_{i=t-t_p}^t$  be a sequence of the 3DoF poses between time $t-t_p$ and time $t$. At each time step $t$, given  $P_{t-t_p:t}$  we predict the user pose $\hat{p}_{t+t_H}$ at time $t + T_H$ such that the distance between the ground truth future pose and the predicted pose is minimized.

The orientation prediction task can be formulated as a sequence modeling problem, in which  Recurrent neural network (RNN) can tackle efficiently. In this work, we utilize Gated Recurrent Units (GRUs) since they have been shown to perform well for smaller datasets \cite{chung_empirical_2014}. For a time horizon $T_H$ and history window $T_P$, the model input  $\mathbf{X} \in \mathbb{R}^{N\times T_p \times D}$,  where $N$ is the number of users and $D$ is the dimension of the input vector, equals 4 for quaternions, is fed into a GRU unit with 128 hidden cells followed another GRU unit of the same size and the output is fed into a linear layer. % as shown in Fig. \ref{fig:gru_architecture}. 

The model output is $\mathbf{y} \in \mathbb{R}^{N\times D}$, where $y_i$ is the prediction for the i$^{\text{th}}$ user at time $t+T_H$ in the future and is trained to minimize the root-mean-square error (RMSE). However, care should be taken when comparing quaternions because, for instance, the distance between $q$ and $-q$ is $2q$ even though they represent the same orientation. Hence, in this work, we measure the angular distance between two quaternions using the angle of rotation $\theta \in [0,\pi]$ needed to transform one orientation into the other and it is calculated as $\theta = 2 \times \cos ^{-1} (|\hat q_0 \cdot \hat q_1|)$, where $\hat q$ is the normalized quaternion, $q_1 \cdot q_2$ is the dot product and $|\cdot|$ is the absolute function.

\subsection{Viewport Streaming}

\begin{figure}[t]%[htbp]
  \begin{center}
    % below the size of the figure has been reduced for example
    \includegraphics[clip,trim=6cm 5cm 5cm 3.5cm,width=0.85\columnwidth]{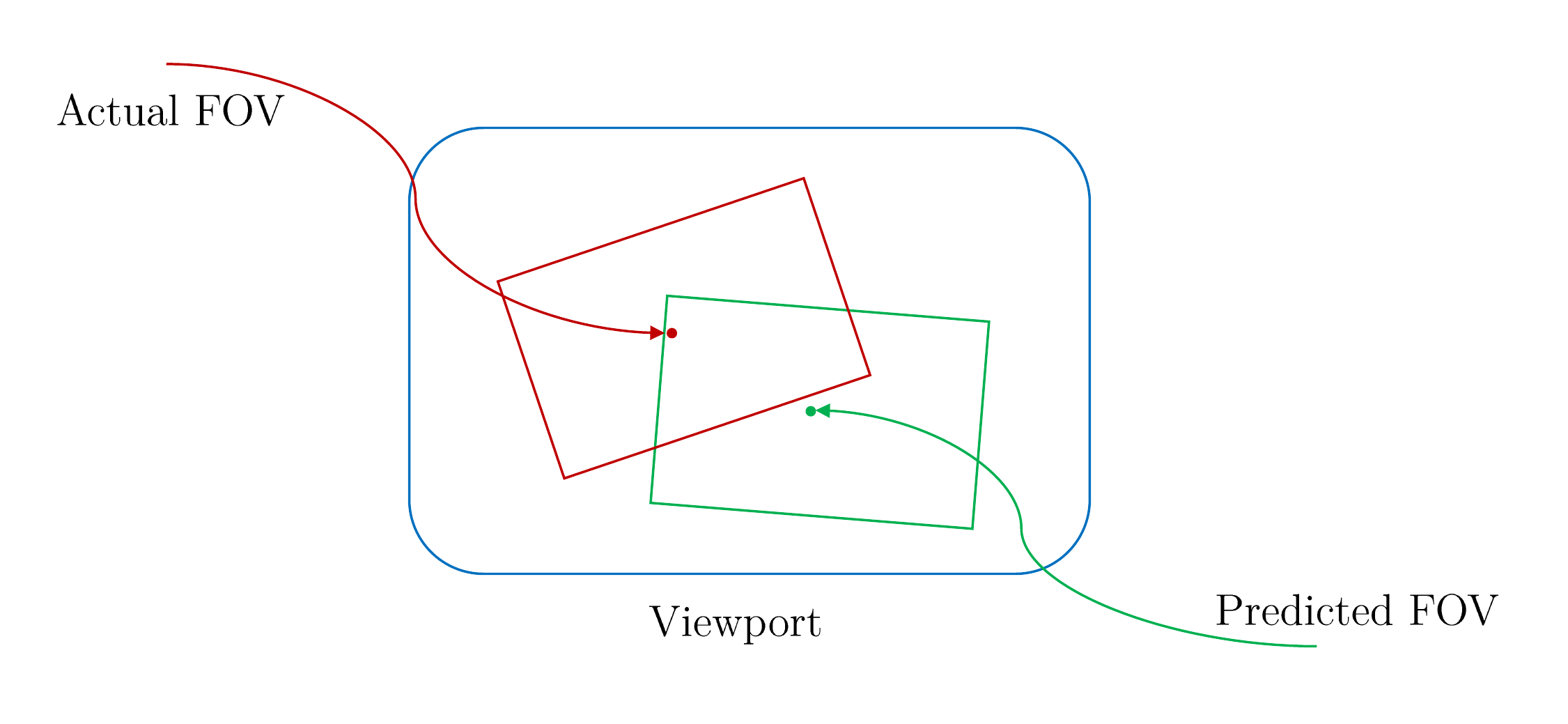}
    \caption[The larger size of the viewport allows the system to tolerate the inaccuracies in predicting the users' orientation.]{The larger size of the viewport allows the system to tolerate the inaccuracies in predicting the users' orientation. Illustration adopted from \cite{bao_motion-prediction-based_2017}.}
    \label{fig:viewport_illustration}
  \end{center}
\end{figure}

We define a viewport-based streaming scheme in which viewports are fixed and overlapping regions that cover the entire sphere.  The size and location of each viewport can be configured based on the capabilities of the rendering server with the constraint that any user's FoV must lie entirely in at least one viewport. Because the viewport size is larger than the users' FoV, which typically only accounts for approximately 1/6 of the entire sphere, this allows the system to tolerate the inaccuracies of the machine learning prediction (see Fig.~\ref{fig:viewport_illustration}) so that for small error values, the predicted orientation is still mapped to a correct viewport that contains the actual orientation.  

A user's FoV may also lie entirely in multiple viewports at the same time. This offers two main advantages. First, since the viewports are predefined, they can be easily pre-rendered and cached at the edge cloud, improving the overall system performance. Second, since the viewports are overlapping, the system can have multiple candidate viewports to choose for a given FoV. The choice of the candidate viewport is controlled by a policy to achieve a certain objective. For instance, one policy is to select the minimum number of viewports that cover a set of users and transmit these via multicast. Another policy would be to select the most popular viewports that can be available in the edge cloud cache.
\subsection{Users-SBS Matching and Scheduling}
The assignment of users to SBS and scheduling of frames can be formulated as a matching game between the SBS and users with active requests, each aiming to minimize latency. This process is repeated at each time instant, hence, a matching procedure is applied to prioritize users based on their requests and the network load.

Let $U = \{u_1, u_2, ..., u_n\}$ be the set of users and $B = \{b_1, b_2, ...., b_m\}$ be the set of SBS. A matching $\mathcal{S}$ is a set of ordered pairs from the set of all ordered pairs $U \times B$ such that each member of $U$ and $B$ appears at most once in any pair in $\mathcal{S}$. Furthermore, each member in $U$ and $B$ has a preference list that ranks members of the opposite set, for example, a user $u$ is said to prefer $b$ to $b'$ if $u$ ranks $b$ higher than $b'$ in $u$ preference list. A pair $(u, b)$ is blocking with respect to a matching $\mathcal{S}$ if $(u, b)$ does not belong to $\mathcal{S}$ and each of $u$ and $b$ ranks the other higher than their partner in $\mathcal{S}$. A matching that contains no blocking pairs is a stable matching. Gale and Shapely \cite{gale_college_1962} proved that each matching problem has a stable matching. They introduced the Deferred Acceptance (DA) Algorithm which is guaranteed to find a stable matching in polynomial time, with $O(n^2)$ worst case runtime. 

In this work, the user-SBS preference is chosen to minimize latency, which can be approximated as the ratio between the required rate (bits) and the actual service rate (bits/s). In the considered video streaming scenario, the required rate is constant as users are requesting fixed sized viewports. Furthermore, only the estimated service rate can be known before transmission since interference cannot be calculated a priori. Hence, the latency is inversely propositional to the estimated service rate.  Hence, we consider the user-SBS preference for a user $u$ and SBS $b$ as:
\begin{equation}
\text{Pref}_{ub}(t) = \text{BW}_b \log_2(1+\text{SINR}_{bu}(t))
\end{equation}
where  $\text{SINR}_{bu}$ is calculated as defined in Eq. \ref{eqn:sinr} with the interference estimated using an exponential moving average  $\hat{I}_u$ and instantaneous interference $\tilde{I}_u$ at time instant $t-1$ with a learning parameter $\beta$. The instantaneous interference $\tilde{I}_u(t)$ on a user $u$ due to currently transmitting active users is given by:
\begin{equation}
\tilde{I}_u(t) = \sum \limits_{b \in B'} \frac{p_b G_{Rx}^{bu}(t) G_{Tx}^{bu}(t)}{\ell_{ub}(t)}
\end{equation}
where $B'$ is the set of interfering base stations and $\ell_{ub}$ is the path loss, the  interference is then estimated as $I_u$:
\begin{equation}
I_u(t) = \beta \tilde{I}_u (t-1) + (1-\beta) \hat{I}_u(t-1)
\end{equation}

Only users with non-empty queues enter the matching procedure. For a network with $\mathcal{B}$ SBSs and $\mathcal{U}$ users, at each transmission instant, at most $\mathcal{B}$ users can have an active transmission. Once a stable matching is reached, the network prioritizes each user queue according to their deadlines. Each user queue can contain either or all of (1) predicted viewports, with their deadlines being after the time horizon $T_H$. (2) sensory-based viewports, with real-time deadline. The latter are admitted to into the queues in case of a false positive prediction. This process is performed by the scheduler and is outlined in Algorithm \ref{algorithm:user_scheduling}.

\begin{algorithm}[t]
\caption{Scheduling between SBSs and Users}
\label{algorithm:user_scheduling}
\textbf{Stage I - Ready-to-schedule users}

\begin{algorithmic}\footnotesize
\State  Let $\text{vp}_{\texttt{prediction}}(t)$ to be the predicted viewport received  at time $t$.
\State  Let $\text{vp}_{\texttt{real-time}}(t)$ to be the actual viewport  received at time $t$. 
\For{user $u \in \mathcal{U}$} %queue admission
        \State Append  $\text{vp}_{\texttt{prediction}}(t)$ to user queue with deadline $t + T_H$ .
         \If {$\text{vp}_{\texttt{real-time}}(t)$ $\neq$   $\text{vp}_{\texttt{prediction}}(t - T_H)$} 
        \State Append  $\text{vp}_{\texttt{real-time}}(t)$ to user queue with deadline $t$.
        \EndIf 
\EndFor
\State $\mathcal{U}' \gets \{ u \in \mathcal{U} \mid \text{ queue is non empty}\}$  \Comment{eligible set of users}
\end{algorithmic}

\textbf{State II - Matching and scheduling}

\begin{algorithmic}\footnotesize
\State For each user $u \in \mathcal{U}'$ and SBS $b \in \mathcal{B}$, calculate $ \text{Pref}_{ub} (t)$.
\State Initialize the subset of unmatched users $\mathcal{S}_{\mathcal{U}} \subseteq \mathcal{U'}$ so that initially $|\mathcal{S}_{\mathcal{U}}| = |\mathcal{U'}|$.
\State Initialize the subset of unmatched SBSs $\mathcal{S}_{\mathcal{B}} \subseteq \mathcal{B}$ so that initially $|\mathcal{S}_{\mathcal{B}}| = |\mathcal{B}|$.
\State For each user $u \in \mathcal{U}'$, initialize the subset of proposal candidates  $\mathcal{P}_{u} \subseteq \mathcal{B}$ so that initially $|\mathcal{P}_{u}| = |\mathcal{B}|$.
\While {$|\mathcal{S}_{\mathcal{U}}| \neq \emptyset$ and $\forall_{u \in \mathcal{U}'} |\mathcal{P}_{u}| \neq \emptyset$}
\State Pick a random user $u \in \mathcal{S}_{\mathcal{U}}$ such that $\mathcal{P}_{u}$ is not empty.
\State Let $b$ be the highest-ranked SBS in $\mathcal{P}_{u}$.
\If {$b \in \mathcal{S}_{\mathcal{B}}$}  \Comment{$b$ is free} 
\State Match $b$ and $u$ and remove them from $ \mathcal{S}_{\mathcal{U}}$ and $\mathcal{S}_{\mathcal{B}}$ respectively.
\Else  \Comment{$b$ is matched with another user $u'$} 
\If {$ \text{Pref}_{u'b} (t) > \text{Pref}_{ub} (t)$} \Comment {$b$ ranks $u'$ higher than $u$} 
\State Refuse proposal from $u$.
\State Remove $b$ from $\mathcal{P}_{u}$.  \Comment{$b$ is no longer a candidate for $u$} 
\Else 
\State Unmatch $b$ from $u'$ and add $u'$ back to $\mathcal{S}_{\mathcal{U}}$
\State Match $b$ and $u$ and remove $u$ from $ \mathcal{S}_{\mathcal{U}}$
\EndIf
\EndIf
\EndWhile
\State For each user $u \in \mathcal{U}_{M}$, sort user queue in descending order of deadline.

\end{algorithmic}
\end{algorithm}

\subsection{Caching}

In order to estimate the spatial popularity profile of any video, we analyze the historical viewing data to rank the viewports and define the popularity of viewport $\text{vp}_f$ at frame index $f$ as the ratio of the users that requested viewport $\text{vp}_f$ to the total number of users. The $K$ most popular viewports at each frame index are then cached at the edge cloud. This number $K$ is dependent on the storage capacity available at the edge cloud and the viewport configuration used, for example a viewport in a 3x3 configuration is 1.5 larger than a viewport in a 5x5 configuration.

\section{Simulation Results}
\label{chapter:evaluation}

In this section, we evaluate the performance of the proposed system on a public dataset collected from 59 users watching seven 360$\degree$ videos, each about 60 seconds long \cite{corbillon_360-degree_2017}.  

\subsection{Machine Learning Training \& Evaluation}
\label{section:chapter_5_ml_training}

A separate model is trained for each video and the training and validation sets contain 80\% and 20\% of the traces respectively. The models were trained using the Adam optimizer with no weight decay and $(\beta_1=0.9, \beta_2=0.999)$ for the running averages coefficients. The models were trained for 100 epochs and a batch size of 32. This process is repeated using K-fold cross validation (K = 5).

\begin{figure}[b]%[htbp]
    \centering
    \includegraphics[width=0.9\columnwidth]{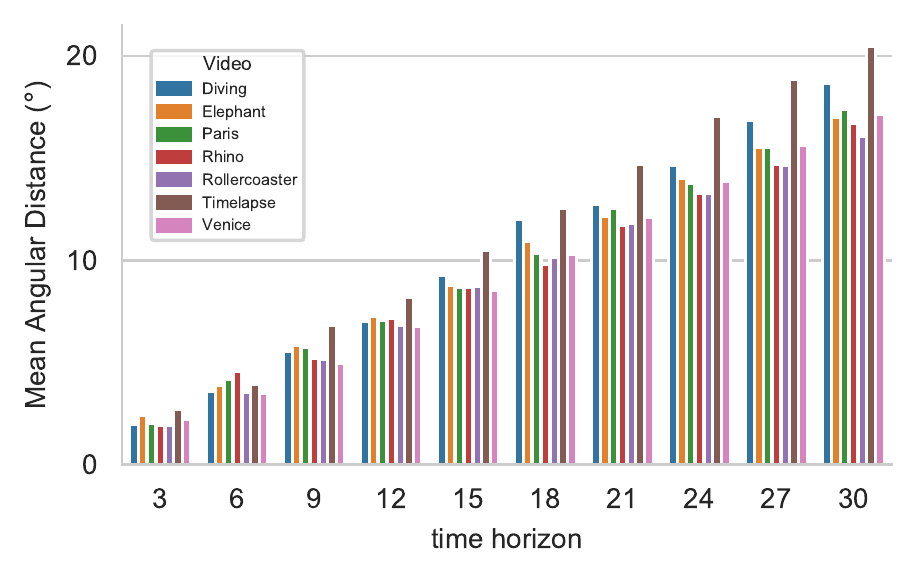}
	\caption{Aggregated Model performance for different videos for time step = 10 frames.}
	\label{fig:ml_imact_of_timestep_by_model}
\end{figure}

The models had a different performance profile among the different videos in the dataset. Overall, there is a rising linear trend in error versus time horizon with the lowest error  achieved in `Rollercoaster` followed by `Rhino` and `Venice` whereas `Timelapse` had the highest error, as shown in Fig.~\ref{fig:ml_imact_of_timestep_by_model}.

The actual and predicted users' orientations are then mapped to the equivalent viewports to be used in the network simulations. In this work, we consider the 3x3 and 5x5 viewport configuration as they provide a good trade-off between the required storage space and overall viewport size.

\subsection{Network Simulations}
\label{section:chapter_5_network_simulations}

We consider a square VR arcade with 100m side length. SBSs are located at the 4 corners of the arcade and users are uniformly distributed throughout the arcade. The pathloss model follows the probabilistic model defined in Section \ref{section:chapter3_network_model} and values are chosen according to 3GPP standard \cite{3gpp_3rd_2019}. The pathloss $\ell_{ub} (d)$ is recalculated every 100ms during the simulation with LOS probability $p(d)$ calculated as:

\begin{equation}
p(d) = \left\{\begin{matrix}
 1.0 & d \leq 5m \\ 
 \exp(-\frac{d-5}{70.8})& 5m < d \leq 49\\ 
 \exp(-\frac{d-49}{211.7}) \times 0.54 & d > 49m
\end{matrix}\right.
\end{equation}

Table \ref{table:simulation_rparameters} provides the set of default parameter values used for the simulations.

\begin{table}[htbp]\footnotesize
\centering
\caption{Simulation Parameters.}
\begin{tabular}{|l|l|}
\hline
\textbf{Parameter} & \textbf{Value}  \\ \hline
Simulation time	& 30 s \\ \hline
Transmission slot duration	& 0.25 ms \\ \hline
Bandwidth	& 0.85 GHz  \\ \hline
mmWave Band	& 28 GHz \\ \hline
Rx beamwidth	& 45$\degree$ \\ \hline
Tx beamwidth	& 90$\degree$ \\ \hline
Noise spectral density	& -174 dBm/Hz \\ \hline
SBS transmit power	& 24 dBm \\ \hline
Users per video 	& 12 \\ \hline
Catalog size	& [3, 4, 5, 6] videos \\ \hline
Time step window	& 10 frames \\ \hline
Prediction horizon	& [5, 10, 15, 20, 25 ,30] frames \\ \hline
Frame duration	& 33 ms \\ \hline
Backhaul latency	& 3 ms \\ \hline
Cache size	& 2 viewports per video frame \\ \hline
\end{tabular}
\label{table:simulation_rparameters}
\end{table}

The following 4 schemes are considered for comparison:

\begin{itemize}[leftmargin=.0cm,itemindent=0.5cm,labelwidth=\itemindent,labelsep=0cm,align=left]
\item \textbf{ML:} Requests are proactively scheduled.
\item \textbf{ML (cache):} Requests are proactively scheduled and popular viewports are cached at the edge cloud.
\item \textbf{NML:} Requests are scheduled in real-time.
\item \textbf{NML (cache):} Requests are scheduled in real-time and popular viewports are cached at the edge cloud.
\end{itemize}

The performance of each scheme is evaluated through average, and 99th percentile delay. Furthermore, transition quality percentage, defined as the ratio of times HD frame delivery failed and HMD had to downgrade to a lower quality frame or interpolate previous frames to the total number of frames, and the HD delivery rate and measured. The metrics are averaged over all users.

\begin{figure*}[htbp]
	\centering  % Remember to centre the figure
	\begin{subfigure}{0.245\textwidth}
	\centering  % Remember to centre the figure
    \includegraphics[clip,trim=2cm 6.7cm 20cm 7cm,width=0.92\textwidth]{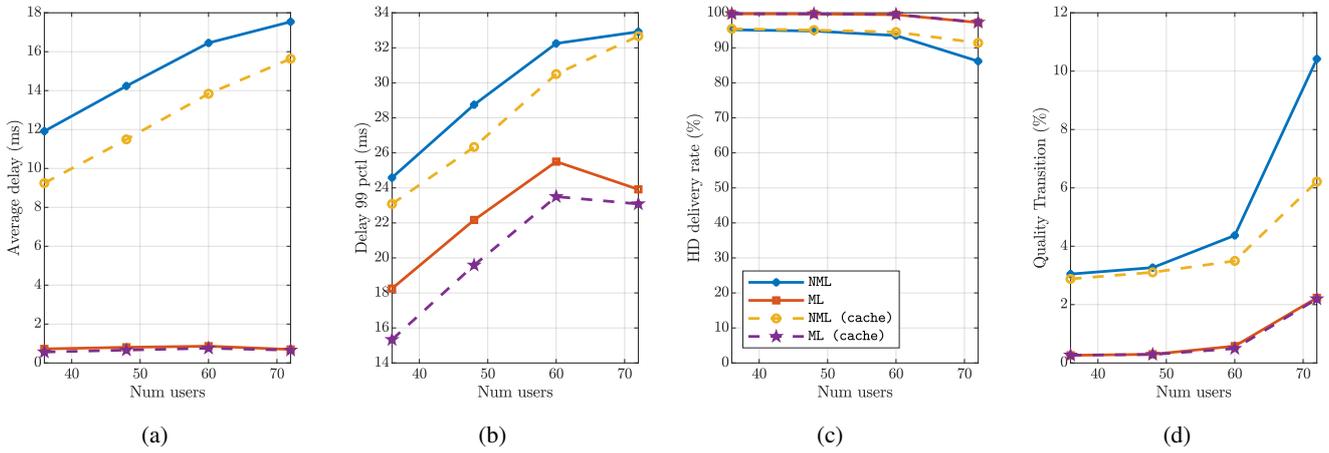}
    \caption{}
    \end{subfigure}%
    \begin{subfigure}{0.245\textwidth}
     \centering
     \includegraphics[clip,trim=8cm 6.7cm 14cm 7cm,width=0.92\textwidth]{images/chapter5_numusers_5x5_10TH.pdf}
	\caption{}
	\end{subfigure}%
	\begin{subfigure}{0.245\textwidth}
     \centering
      \includegraphics[clip,trim=14.15cm 6.7cm 7.85cm 7cm,width=0.92\textwidth]{images/chapter5_numusers_5x5_10TH.pdf}
	\caption{}
	\end{subfigure}
	\begin{subfigure}{0.245\textwidth}
     \centering
     \includegraphics[clip,trim=20.5cm 6.7cm 1.5cm 7cm,width=0.92\textwidth]{images/chapter5_numusers_5x5_10TH.pdf}
	\caption{}
	\end{subfigure}
	\caption[ Delay, HD delivery rate and quality transition versus the number of users in a 5x5 viewport configuration with $T_H$ = 10.]{(a) Average delay, (b) 99th percentile delay, (c) HD delivery rate and (d) Quality transition versus the number of users (with a fixed number of 12 users per video) in a 5x5 viewport configuration with $T_H$ = 10.}
	\label{fig:numusers_5x5_10TH}
\end{figure*}

\paragraph{Impact of Network Load}
The effect of the network load is analyzed by simulating the network under an increasing number of users and videos. Fig. \ref{fig:numusers_5x5_10TH} shows that the proposed schemes achieve lower latency and maintain close to 100\% HD rate up to 60 users. Furthermore, caching helps significantly reduce the rising slope of the 99th percentile delay, this is observed in the different schemes, with more prominence in non-ML schemes. Intuitively, for real-time requests, fetching content from the remote cloud is on the critical path and any speedup in fetching improves the overall latency. However, in ML schemes and with the use of proactive scheduling, fetching content is no longer on the critical path as the network has $T_H$ frames to deliver the content.

\paragraph{Impact of Prediction Horizon}

\begin{figure*}[htbp]
	\centering  % Remember to centre the figure
	\begin{subfigure}{0.245\textwidth}
	\centering  % Remember to centre the figure
    \includegraphics[clip,trim=2cm 6.7cm 20cm 7cm,width=0.92\textwidth]{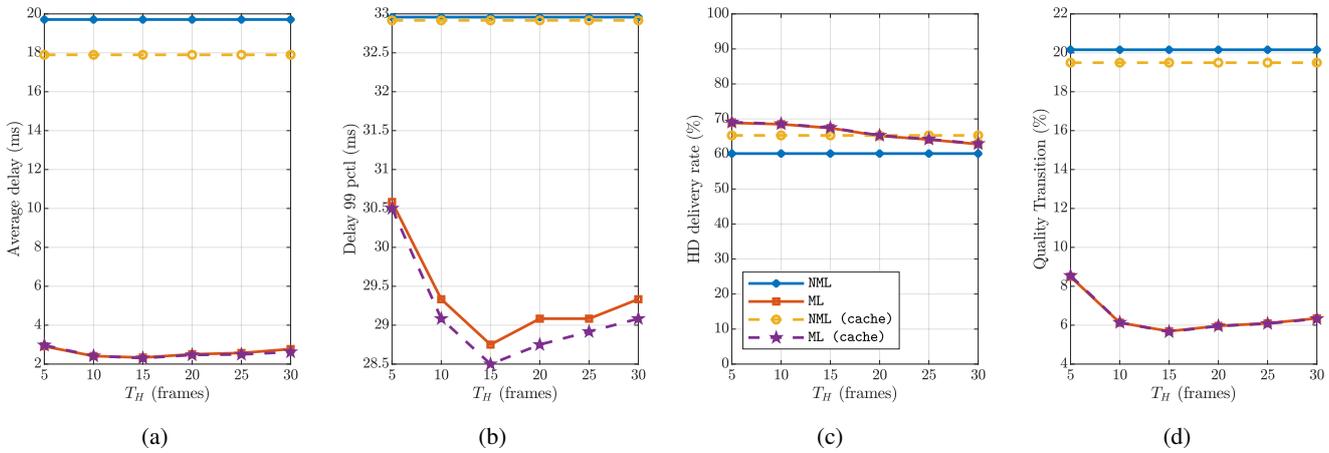}
    \caption{}
    \end{subfigure}%
    \begin{subfigure}{0.245\textwidth}
     \centering
     \includegraphics[clip,trim=8cm 6.7cm 14cm 7cm,width=0.92\textwidth]{images/chapter5_timehorizon_3x3_72users.pdf}
	\caption{}
	\end{subfigure}%
	\begin{subfigure}{0.245\textwidth}
     \centering
      \includegraphics[clip,trim=14.15cm 6.7cm 7.85cm 7cm,width=0.92\textwidth]{images/chapter5_timehorizon_3x3_72users.pdf}
	\caption{}
	\end{subfigure}
	\begin{subfigure}{0.245\textwidth}
     \centering
     \includegraphics[clip,trim=20.5cm 6.7cm 1.5cm 7cm,width=0.92\textwidth]{images/chapter5_timehorizon_3x3_72users.pdf}
	\caption{}
	\end{subfigure}
	\caption[ Delay, HD delivery rate and quality transition versus the prediction horizon (in frames) in a 3x3 viewport configuration for 72 users.]{(a) Average delay, (b) 99th percentile delay, (c) HD delivery rate and (d) Quality transition versus the prediction horizon (in frames) in a 3x3 viewport configuration for 72 users.}
	\label{fig:timehorizon_3x3_72users}
	\vspace{-0.4cm}
\end{figure*}

A longer prediction horizon $T_H$ allows the scheduler more leeway in scheduling future frames, however, the prediction error monotonically increases with the prediction horizon, as discussed in Section \ref{section:chapter_5_ml_training}, resulting in more data to be transmitted in real-time. Hence, the baseline methods remain constant as they do not proactive schedule frames whereas in the proposed methods the delay is affected by (1) the network load and ability to meet the current demand and (2) the prediction error causing misses and for frames to be scheduled in real-time. Fig. \ref{fig:timehorizon_3x3_72users}  shows the performance using 72 users for a 3x3 viewport configuration. In this heavily loaded scenario,  the delay exhibits a U-shaped performance where the delay initially decreases due to having more scheduling time, then increases again due to high miss rate and the need to schedule more frames in real-time.

As such, even with non-perfect predictions, the use of proactive scheduling has significantly improved the latency and delivery rate. Proactive scheduling is especially useful in fluctuating channels; by transmitting viewports a few frames in advance, a transient drop in users' rate budget would not result in an outage.

\section{Conclusion}
\label{chapter:conclusion}

This paper has presented a solution for proactive VR viewport streaming in wireless network. The proposed solution minimizes latency and meets the high bandwidth and low latency constraints by combining  user orientation prediction, caching, and proactive delivery of user's own viewport of the 360$\degree$ video frame. Machine learning model based on recurrent neural networks has been developed and has been shown to predict user orientation at a future time horizon based on previous motion history.  Subsequently, Deferred Acceptance matching algorithm has then been used to match users and base stations with the preferences defined to minimize estimated latency. The combination of prediction and matching for proactive streaming is shown to significantly minimize latency and maximize delivery rate. Caching has been leveraged to store popular video viewports in edge servers close to users which is shown to minimize the latency, especially in real-time requests. \vspace{-0.2cm}

\bibliographystyle{IEEEtran}
\bibliography{conference.bib}

\end{document}